\documentclass[aps,prb,amsmath,amssymb,twocolumn,floatfix,10pt,showpacs]{revtex4-1}
\usepackage{graphicx}
\usepackage{hyperref}
\usepackage{color}

\newcommand{\ve}{\mathbf}

\renewcommand{\vec}[1]{\mathbf{#1}}

\global\long\def\ket#1{\left|#1\right>}

\global\long\def\komm#1#2{\left[#1,#2\right]}

\begin{document}

\title{Effective spin theories for edge magnetism in graphene zigzag ribbons}
\author{Cornelie Koop and Manuel J. Schmidt}
\affiliation{Institut f\"ur Theoretische Festk\"orperphysik, RWTH Aachen University, 52056 Aachen, Germany}
\date{\today}

% 73.43.Nq 	Quantum phase transitions
% 73.20.-r Electron states at surfaces and interfaces

\begin{abstract}
We report a thorough study of the reducibility of edge correlation effects in graphene to 
much-simplified effective models for the edge states. The latter have been used before in specially 
tailored geometries. By a systematic investigation of corrections due to the bulk states in second 
order perturbation theory, we show that the reduction to pure edge state models is well-justified in 
general geometries. The framework of reduced models discussed here enables the study of 
non-mean-field correlation physics for system sizes far beyond the reach of conventional 
methods, such as, e.g., quantum Monte-Carlo.
\end{abstract}

\maketitle

\section{Introduction}

Due to its vanishing density of states at the charge neutrality point, bulk graphene,
\cite{geim_rise_of_graphene_2007,graphene_rmp_2009} a two-dimensional hexagonal lattice of 
carbon atoms, shows usually no features of strong magnetic correlations. But disturbances of the 
hexagonal lattice, such as edges or vacancies, modify the {\it local} density of states in that they 
generate peaks close to the Fermi level. This presence of states localized at lattice disturbances is 
experimentally and theoretically well established. Edge states are routinely observed in scanning 
tunneling spectroscopy\cite{kobayashi_2005,niimi_2005,niimi_2006,tao_2011} and consistently 
appear in theoretical investigations.
\cite{fujita_1996,nakada_1996,wakabayashi_analytical_edge_states_2010} The magnetic 
properties of these states, however, are not as obvious. Early theoretical mean-field studies of the 
Hubbard model on the honeycomb lattice\cite{fujita_1996,wakabayashi_edge_magnetism_1998} 
predicted finite spin polarizations at zigzag edges of graphene, a phenomenon known as edge 
magnetism. These studies were later shown to be consistent with density functional theory calculations,
\cite{lee_edge_magnetism_2005,son_abinitio_prl_2006,yazyev_mag_corr_2008} but spin 
polarizations have never been observed experimentally. 

All theoretical investigations the previous paragraph referred to had one thing in common: they all 
used some sort of self-consistent field (SCF) approach in order to describe the magnetic 
properties, with the local spin polarization being the self-consistent field. It is worth noting, however, that the 
presence of such a polarization is in contradiction with Lieb's theorem,\cite{lieb_theorem} even 
though some studies explicitly say otherwise. In short: Lieb's theorem states that a graphene 
nanoribbon has a spin-singlet ground state. SCF theories  predict spin 
polarizations with opposite directions at opposite edges. The vanishing {\it total} spin polarization 
has erroneously been said to indicate consistence with Lieb's theorem, but in fact only a vanishing 
{\it local} spin polarization is consistent with a singlet ground state and thereby with Lieb's 
theorem. Being overly pointed, one might say that edge magnetism is only a mean-field artifact. 
And defining edge magnetism only as the presence of a local spin polarization in the ground state of an 
isolated (possibly infinitely long) graphene nanoribbon, it probably {\it is} an artifact. It turns out, 
however, that with this very restrictive definition of edge magnetism, one cannot resolve the 
intriguing magnetic correlation phenomena that can be studied only with non-mean-field methods, 
such as quantum Monte-Carlo\cite{feldner_prb_2010,feldner_prl_2011,golor_chiral_qmc_2013} 
(QMC) or the density matrix renormalization group\cite{hikihara_2003} (DMRG).

In a recent study, the nature of edge magnetism has been addressed from a broader perspective of 
dynamics and open systems.\cite{golor_quantum_edge_magnetism} No mean-field approximation 
has been used and the spin fluctuations, which are essential for the ground state being a singlet, 
have been respected properly. The central novelty of this perspective is that, although the ground 
state of an isolated ribbon shows indeed no static polarization, the environment-induced 
decoherence may stabilize a finite non-equilibrium spin polarization. It has been argued with the 
quantum Zeno effect: the continuous observation by the environment stabilizes the 
finite-spin-polarization state, which, in an isolated ribbon, would decay after a short time. Thus, 
whether edge magnetism manifests itself as a spin polarization or not, depends on the 
environment, which tends to destroy quantum correlations and makes systems behave more 
classical.

The study mentioned in the last paragraph was only possible because of the previous development 
of efficient low-energy theories of the electronic correlations at graphene edges.
\cite{tem_schmidt_loss_2010,luitz_ed_2011,schmidt_eff_vs_qmc_2013} These enable the 
investigation of very large ribbons and even the study of time-dependent phenomena. The latter is 
still approximate, but, unlike SCF theories, well controlled. In these effective theories the electronic 
correlation physics is completely described by a quantum spin model, which is derived from the 
parent Hamiltonian (Hubbard model on the honeycomb lattice) by a sequence of controlled approximations. Unfortunately, rather restrictive 
edge geometries (see Fig. 1 in Ref. \onlinecite{golor_quantum_edge_magnetism}) have been 
necessary in order to maintain the high quality of the effective spin theory. In fact, the chiral edges 
considered in Ref. \onlinecite{golor_quantum_edge_magnetism} were designed such that the edge states could be transformed to a basis where each zigzag 
segment contains exactly one strongly localized electronic state, overlapping weakly with all other 
states. Moreover, the bulk states were completely neglected. The actual impact of the bulk states, 
especially if the edge states are not as nicely localized as in the geometry used in Ref. 
\onlinecite{golor_quantum_edge_magnetism}, is not clear.

In this work we study those corrections to the low-energy sector from the bulk states, which have 
been neglected before. For this, we directly work in the worst-case scenario, namely a zigzag 
nanoribbon, in which the overlap between neighboring Wannier edge states is maximal. The 
coupling between these localized edge states and the bulk states is mediated by the Hubbard 
interaction, which is in principle rather strong. With a proper basis choice, however, the strong and 
weak parts of the interaction may be  separated, and it turns out that the coupling between 
edge- and bulk states is mediated exclusively by the weak parts. Thus, we may treat them 
perturbatively. We employ a second-order Schrieffer-Wolff transformation, the result of which is a 
rather intricate effective Hamiltonian with a huge number of terms describing the bulk-mediated 
coupling between edge states. The derivation and analysis of these terms and in particular the 
identification of the most important ones is the central goal of this paper.

\begin{figure}
\centering

\includegraphics[width=\linewidth]{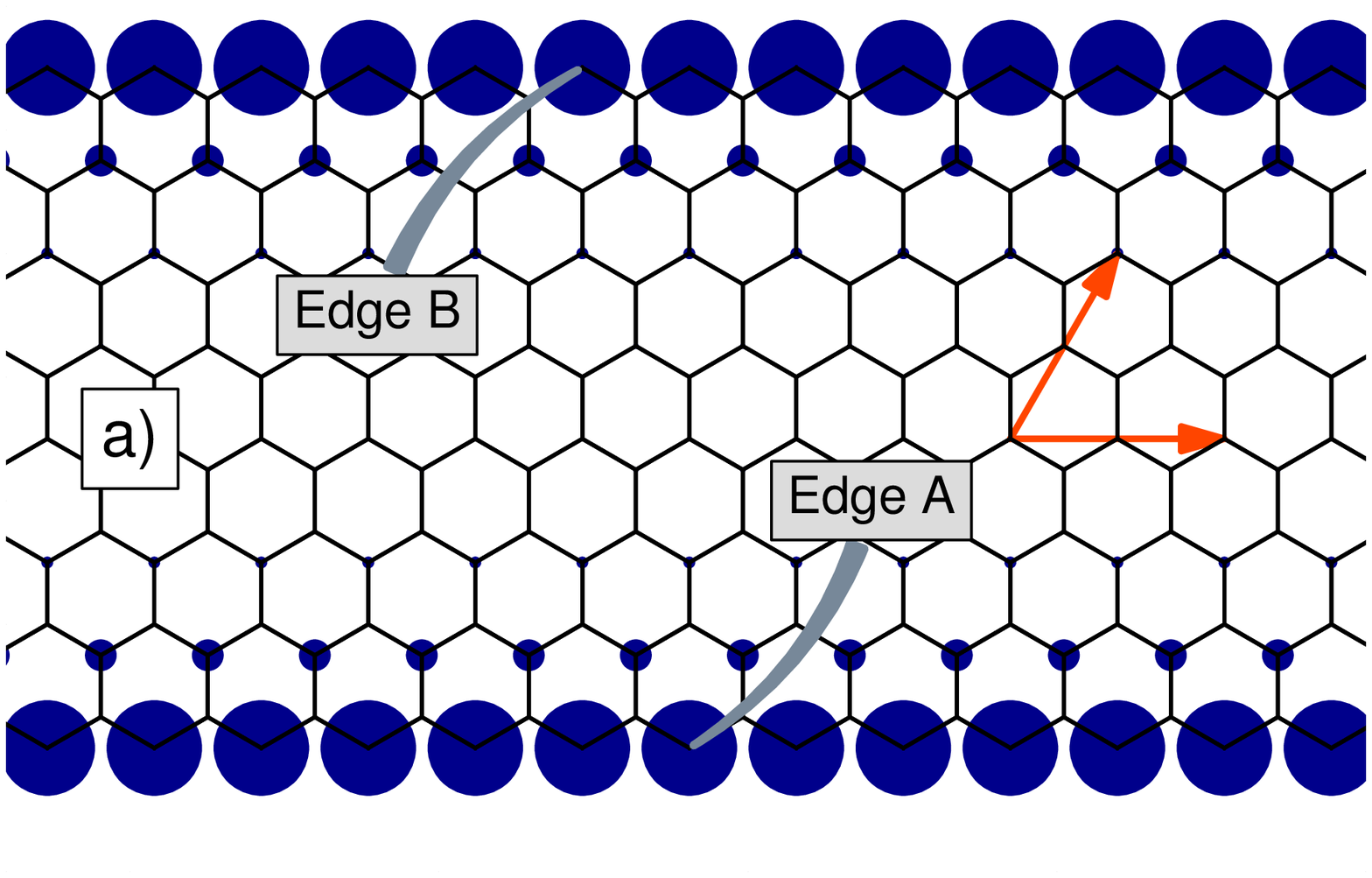} \\%
\includegraphics[width=\linewidth]{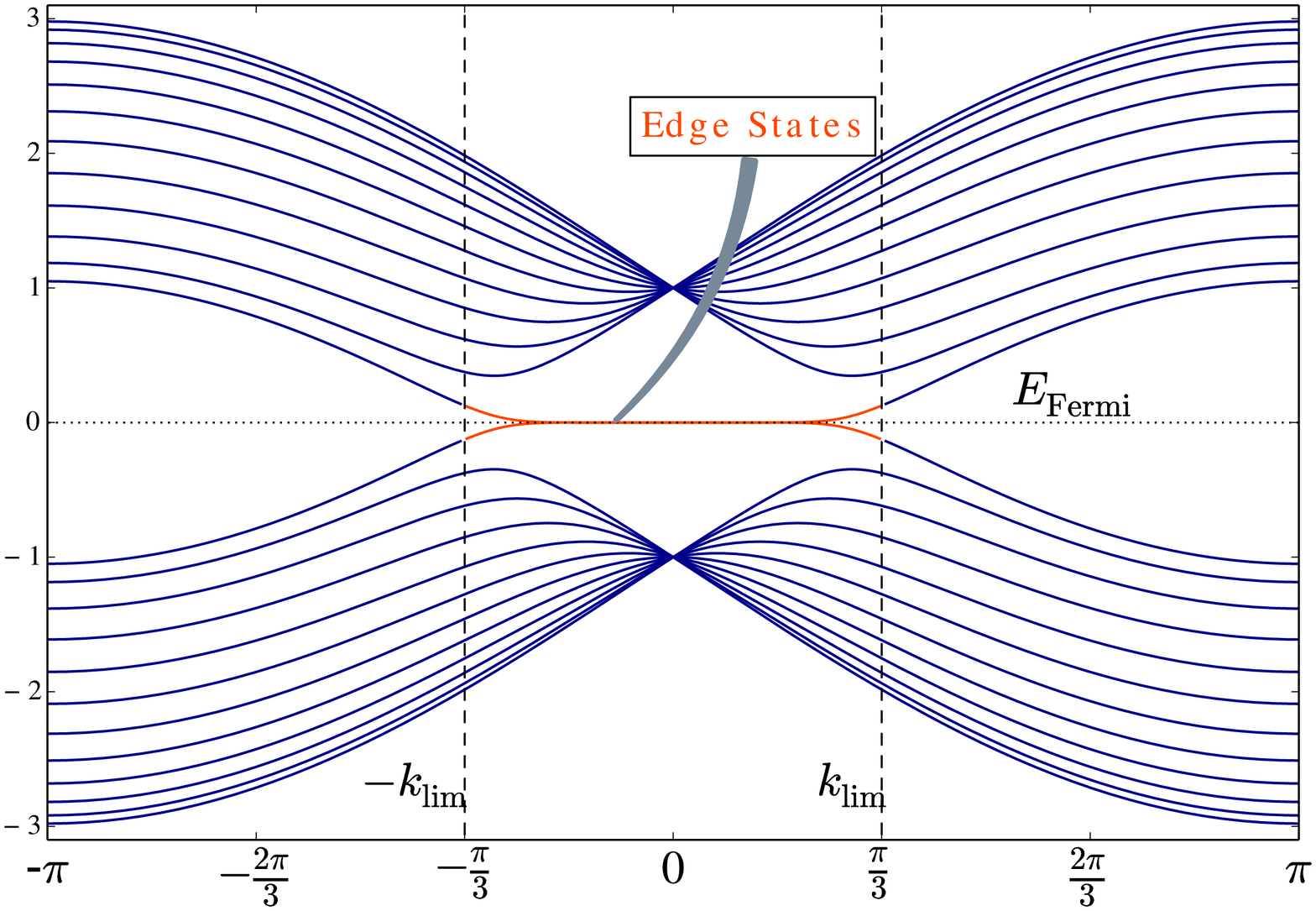}%
\caption{(Color online): a) Detail of a zigzag nanoribbon for $M=10, N=20$. Periodic boundary conditions apply in x-direction. 
The circles show a typical edge state wave function $\psi_{k+}$ and have a radius $r \propto \left|\psi_{k+}(i)\right|$. The wavevector does not include the second basis switch, 
so this is actually an edge eigenstate of the original Hamiltonian Eq. \eqref{H0decoupled}. The weight is equally distributed on both edges of the ribbon.
The (orange) vectors indicate the basic lattice vectors, doubled in length for better visibility. In our basis, the coordinate $n$ is counted in direction parallel to the zigzag edges, and 
$m$ along the second basis vector.
b) Band structure of a ribbon with $M=12, N=200$. The red part of the flat bands belongs to the edge states with $|k|\leq \pi/3$.}%
\label{fig_lattice_bandstructure}
\end{figure}

\section{First order theory\label{sect_first_order}}

\subsection{Hamiltonian and basis choice}

The geometry of the nanoribbon we are interested in is shown in Fig. 
\ref{fig_lattice_bandstructure}. Our theoretical description starts from
the
 Hubbard Hamiltonian on a honeycomb lattice
\begin{equation} \label{H_original}
H= H_{0}+H_{U}=-t\sum_{\left<i,j\right>, \tau} c_{i\tau}^{\dagger}c_{j \tau}+U\sum_{i} : n_{i\uparrow} : :n_{i \downarrow}:\text{,}
\end{equation}
where $c_{i\tau}$ annihilates an electron with spin $\tau=\uparrow,\downarrow$ at lattice
site $i$ and $n_{i\tau} = c^\dagger_{i\tau}c_{i\tau}$. Furthermore, we define the normal order of an operator $:A: = A - \langle A\rangle_0$, with $\langle\cdot\rangle_0$ the average with respect to the non-interacting ground state. $\left<i,j\right>$  is the sum over nearest neighbors.
Throughout this work we assume charge neutrality, i.e., the Fermi level is at zero energy.
The transformation into the eigenbasis of the non-interacting term $H_0$ leads to a complete 
decoupling of the edge and bulk states in $H_0$.
A convenient but somewhat superficial distinction between edge and bulk states can be made on 
the basis of their energies $\epsilon_\mu$: Those states $\mu$ with $\epsilon_\mu$ very close to 
the Fermi level are the edge states. States with higher energies are bulk states. This issue of 
distinguishing edge and bulk states has been discussed in Ref. 
\onlinecite{schmidt_eff_vs_qmc_2013} and will be discussed below. For the moment we define the 
edge states for a given momentum $k$ along the ribbon as the two 
eigenstates of $H_0$ with the lowest absolute energy in the momentum 
interval $[-\pi/3,\pi/3]$ (see Fig. \ref{fig_lattice_bandstructure}). 
Defining $\mathcal E$ ($\mathcal B$) as the set of edge (bulk) states, the hopping Hamiltonian can be written as
\begin{align} \label{H0decoupled}
H_{0} &=  \sum_{\mu \in \mathcal{E},\tau}\epsilon^{e}_{\mu} e_{\mu\tau}^{\dagger} e_{\mu\tau}+\sum_{\mu\in \mathcal{B},\tau} \varepsilon^{b}_{\mu} b_{\mu\tau}^{\dagger}b_{\mu\tau} \\ 
&= H_{0,\rm edge} + H_{0,\rm bulk},
\end{align}
with the bulk and edge state annihilation operators, $b_{\mu\tau}$ and 
$e_{\mu\tau}$, respectively. 
$\mu$ is a collective index, containing momentum ($k$) and subband information.
In the eigenbasis of $H_0$, the interacting part of the Hamiltonian ($H_U$) takes the form 
\begin{equation}\label{HUgeneral}
H_{U}=U\sum_{1234}\Gamma_{\lambda\mu\nu\pi}:d_{\lambda\uparrow}^{\dagger}d_{\mu\uparrow}d_{\nu\downarrow}^{\dagger}d_{\pi\downarrow}:\text{ ,}
\end{equation}
where the operators $d^{(\dagger)}$ can be either edge or bulk operators.
The interaction vertex $\Gamma_{\lambda\mu\nu\pi}$ can be written in terms of the wave functions $\phi_\mu(i)$ of the $H_0$ 
eigenstates
\begin{equation}
 \Gamma_{\lambda\mu\nu\pi}=\sum_{i}\phi_{\lambda}^{*}(i)\phi_{\mu}(i)\phi_{\nu}^{*}(i)\phi_{\pi}(i).\label{interaction_vertex}
\end{equation} 
For the sake of better readability we will sometimes add upper indices b,e to $\Gamma$. These 
indicate whether the lower indices correspond to bulk or edge state operators (e.g. 
$\Gamma^{\rm{ebeb}}_{1234}$ means that indices 1 and 3 correspond to edge states). 

A closer inspection of the edge states $e_{\mu\tau}$ shows that the index $\mu=(k,\pm)$ is 
composed of the momentum along the ribbon $k$ and an additional binary index $\pm$ indicating 
the sign of the edge state energy. The corresponding wave functions $\phi_{k\pm}(i)$ are 
distributed over both edges. For what follows, however, it is more convenient to work in an edge 
state basis where each state is localized at one edge. This can be accomplished by a second basis 
transformation within the edge state subspace $\phi_{k, \rm{A/B}}=\left(\phi_{k+}\pm\phi_{k-} \right)/
\sqrt2$ or, equivalently, $e_{k,\rm A/B,\tau} = (e_{k+,\tau} \pm e_{k-,\tau})/\sqrt2$.
These new states are not eigenstates of $H_0$ [Eq. \eqref{H0decoupled}], but their amplitudes have
non-vanishing values only in the vicinity of one edge. In addition, the amplitudes are non-zero only on the sublattice the edge terminates on, which is why the label A/B indicates edge and sublattice simultaneously. The edge state part of $H_0$ thus reads
\begin{equation}
H_{0,\rm edge} = \sum_{\tau,k} t(k) e^\dagger_{k A} e_{kB} + H.c.,\label{H0edge}
\end{equation}
where the $k$-dependent inter-edge hybridization
\begin{equation} \label{tStarK}
t(k) = \epsilon^{e}_{k+}
\end{equation}
 is equal to the edge state energy in Eq. \eqref{H0decoupled}.

\subsection{Effective interactions}

As far as $H_0$ is concerned, the bulk states are completely independent of the edge states, which is a simple 
consequence of their definition as eigenstates of $H_0$. 
The electron-electron interaction $H_U$, however, contains terms which couple edge and bulk states. 
This can be seen best in Eq. \eqref{HUgeneral}, where the operators $d$ may be edge- or bulk operators in arbitrary combinations. 
Thus, there will be mixed terms containing $0<n<4$ edge- and $4-n$ bulk operators.
We will collect all those terms into the Hamiltonian $H_{\rm int}$, which describes the bulk-edge interaction.
It was noted before (see, e.g., Ref. \onlinecite{tem_schmidt_loss_2010}) that $H_{\rm int}$ is generally small so that a projection of $H_U$ onto
the subspace spanned by the edge states only gives rise to a relatively good approximation of the edge state physics. 
The corrections due to $H_{\rm int}$ can thus be added within perturbation theory. 
It is then customary to introduce a small parameter $\lambda$ with
\begin{equation}
H_{\rm int}\propto \lambda,
\end{equation}
which helps us to keep track of the order in which $H_{\rm int}$ enters.

The investigation of this coupling is the main concern of this work. 
But in this section we first review the consequences of $H_U$ for the edge states if $H_{\rm int}$ is treated to first order. To be definite, we clearly state the approximations used:
\begin{itemize}
 \item Terms that consist only of bulk operators are not regarded, because these contribute to the edge state physics only in higher orders of the edge-bulk interaction.
 \item In order to effectively eliminate the bulk states from our theory, we restrict the Fock space to those states in which all bulk states with energies below the Fermi level (i.e., negative energies) are occupied, while those with positive energies are empty. As a consequence, all terms in Eq. (\ref{HUgeneral}) in which the bulk operators do not appear in creator-annihilator pairs are dropped, because they map a state out of the restricted Fock space.
 \item Finally, the remaining bulk operators are averaged with respect to the Fermi sea, i.e.,
 \begin{equation}
 b_{1\tau}^{\dagger}b_{2\tau'} \;\;\rightarrow\;\; \delta_{12}\delta_{\tau\tau'}\Theta\left(-\varepsilon^{b}_{1}\right).\label{bulk_operator_avg}
 \end{equation}
\end{itemize}
With these approximations the effective edge state interactions can be written as
\begin{equation}\label{HUeeee}
H_{U} \approx U\sum_{\lambda\mu\nu\pi}\Gamma_{\lambda\mu\nu\pi}:e_{\lambda \uparrow}^{\dagger}e_{\mu \uparrow}::e_{\nu \downarrow}^{\dagger}e_{\pi \downarrow}:.
\end{equation}
Note that in order to derive Eq. (\ref{HUeeee}) it is not sufficient to just drop all terms with bulk operators from Eq. (\ref{HUgeneral}). It is essential to perform the bulk state average properly, as described above. Additionally employing particle-hole symmetry of $H_0$, however, the first order of $H_{\rm int}$ can be absorbed completely into the normal ordering of the edge state operators in Eq. (\ref{HUeeee}). Note also that Eq. (\ref{HUeeee}) is valid for any choice of basis in the edge state subspace.

\begin{figure}
\centering
\includegraphics[width=\linewidth]{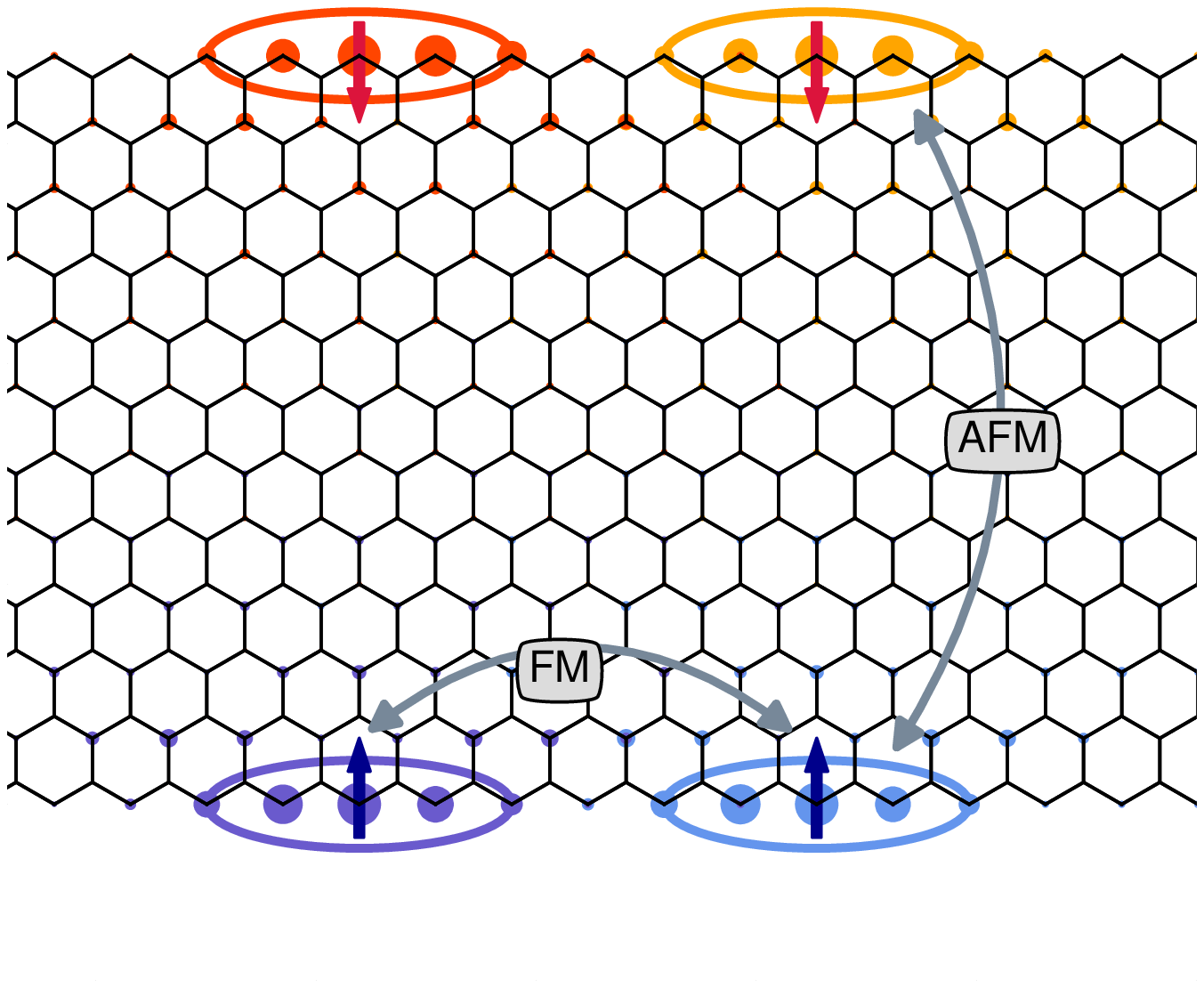} %
\caption{(Color online) The Wannier states lie at either edge of the ribbon and are also localized in x-direction.
In this sketch, the filled circles represent the absolute value of the actual Wannier wavefunctions, while the oval lines are just guides to the eye. 
Please note that the colors belong to next-nearest neighboring states, and with one state in between the depicted ones at each edge 
there is a considerable overlap along the edge. Due to the strong effective onsite repulsion $U \Gamma^{eeee}_{xxxx}$  we can go over to a basis with each state singly occupied.
The ferromagnetic intra-edge interaction couples sites of the same ribbon side, 
while the antiferromagnetic inter-edge term connects edge states on opposite edges.}%
\label{fig_lattice_wannier}
\end{figure}

Solving this approximate fermionic edge state theory is still relatively difficult. In Ref. 
\onlinecite{schmidt_eff_vs_qmc_2013} this has been done with exact diagonalization, and it was 
shown that already at this level, the reduced theory can compete with QMC simulations. But the 
actual strength of this reduction becomes apparent only after another approximation step. For this 
one performs a transformation to a Wannier edge state basis (for details, see Sec. 
\ref{sect_wannier_cutoff}) in which the edge states are not only localized at one individual edge, 
but have also a restricted extent along this edge. In this basis it becomes apparent that, due to the 
structure of $H_U$ [Eq. (\ref{HUeeee})], each Wannier edge state hosts one electron and the 
residual dynamics is well approximated by a pure spin model in which spins at the same (opposite) 
edge are coupled ferromagnetically (antiferromagnetically). 
Figure \ref{fig_lattice_wannier} shows the wave functions of four typical Wannier edge states and 
indicates their spin couplings.
The latter are directly related to the vertex functions $\Gamma$ in Eq. (\ref{HUeeee}) 
and the inter-edge hybridization $t_{xx'}$ in the Wannier basis (details can be found in Sec. \ref{sect_wannier_cutoff}). In particular, the 
ferromagnetic intra-edge coupling between two spins in Wannier sites $x$ and $x'$ is given by\footnote{In this work we will use spin operators $\ve S$ with eigenvalues $\pm1/2$.}
\begin{equation}\label{J0_FM}
J_{xx'}^{\rm{FM}} =2U\Gamma_{xxx'x'},
\end{equation}
while the antiferromagnetic inter-edge coupling reads
\begin{equation}%\label{J0_AFM}
J_{xx'}^{\rm{AFM}}=\frac{4\left(t_{xx'}\right)^{2}}{U\Gamma_{xxxx}}. \label{first_order_afm}
\end{equation}

It is important to note here that the edge states localized at one edge only are also sublattice polarized. 
In combination with the interaction vertex [Eq. (\ref{interaction_vertex})] being a sum over real-space wave function products, 
this results in a kind of selection rule, namely that there are no terms in Eq. (\ref{HUeeee}) coupling different edges. 
Instead, the inter-edge coupling is mediated solely by the one-particle inter-edge hybridization $t_{xx'}$. 

In the remainder of this paper we will extend this model by additionally investigating
second order contributions originating from the coupling  $H_{\rm int}$ of edge and bulk states. As above, we will do this in two steps.
In Section \ref{sect_ferm_bulk_corr} we will derive the corrections to the fermionic theory [Eqs. (\ref{H0edge}) and (\ref{HUeeee})] and in Sec. \ref{sec:SpinModel} we will study the consequences of these corrections for the spin theory.
 
\section{Fermionic bulk corrections\label{sect_ferm_bulk_corr}}

The effect of the bulk states on the edge state interaction is approximated up to second order by means of a Schrieffer-Wolff
transformation. 
For this, we consider the free part of the bulk Hamiltonian $H_{0,\rm bulk}$
as the complete free Hamiltonian and the interaction between
edge and bulk states $H_{\rm int}$ as its perturbation. The free part
of the edge Hamiltonian is kept separately, which means that we neglect
the time-dependence of the edge states. This constitutes a good approximation
for ribbons which are sufficiently broad so that the inter-edge hybridization $t(k)$ is small compared to the typical bulk state energies $\epsilon^{(b)}_{\mu}$. Formally, the second-order Schrieffer-Wolff correction can be written as
\begin{equation}\label{Hsw} 
H_{\text{SW}}=-\lim_{\eta\rightarrow 0}\frac{i}{2}\int_{0}^{\infty}dt\ e^{-\eta t}\komm{H_{\text{int}}(t)}{H_{\text{int}}},
\end{equation}
where $H_{\rm int}(t) = \exp(i t H_{0,\rm bulk} ) H_{\rm int} \exp(-i t H_{0,\rm bulk} )$.
As before, the bulk operators are eliminated by appropriate averaging [Eq. (\ref{bulk_operator_avg})].

After some tedious but straightforward calculations we obtain the complete fermionic expression of the second order Schrieffer-Wolff terms
\begin{equation}
H_{\rm SW} = H^{2p}_{\rm SW} + H^{1p}_{\rm SW} + H^{3p}_{\rm SW} \label{HSW_terms},
\end{equation}
where the individual constituents are effective two-particle, one-particle and three-particle interactions, respectively.
In its totality, this correction seems very involved. Luckily, however, many terms are either small from the beginning, or are suppressed by the large zero- and first order interaction terms in Eq. (\ref{HUeeee}). As a consequence, the relevant bulk corrections can be understood easily in the end. We now discuss each type of SW correction separately.

\subsection{Two-Particle terms}
\begin{figure}[htb] 
\centering
\includegraphics[width= 0.45\linewidth]{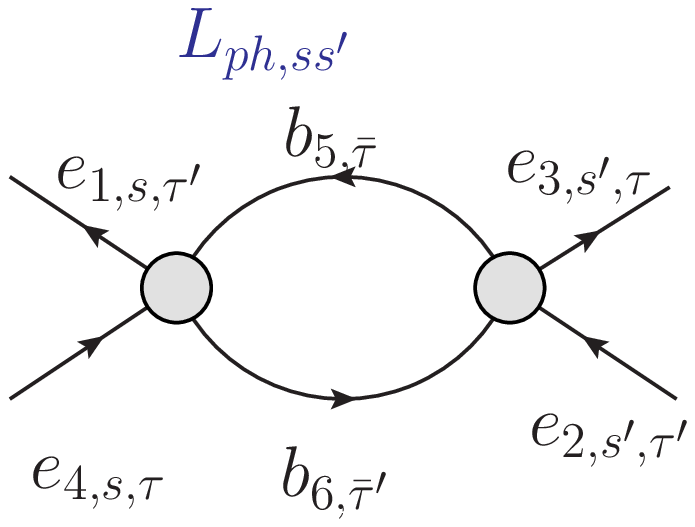}% 
\includegraphics[width= 0.45\linewidth]{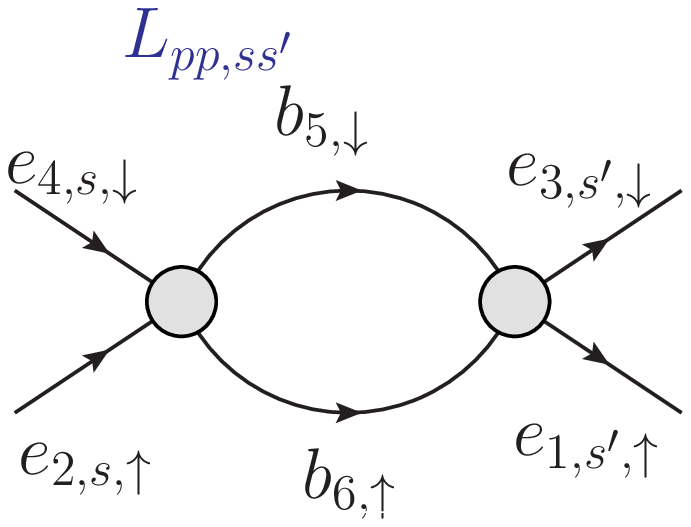} \\ %
and                                                                    \\ %
\includegraphics[width= 0.45\linewidth]{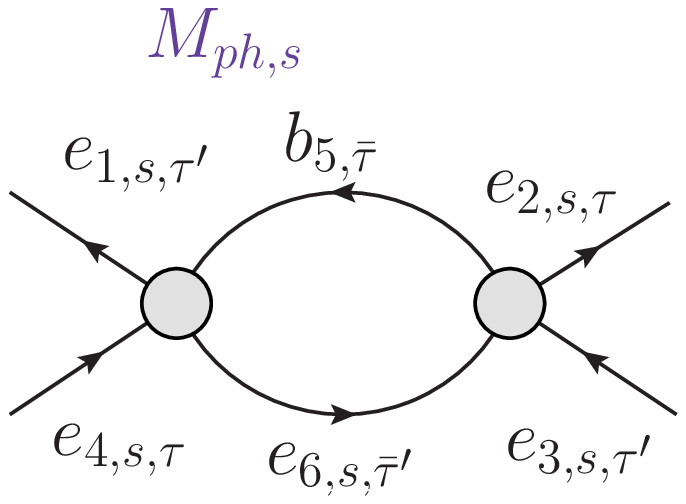}% 
\includegraphics[width= 0.45\linewidth]{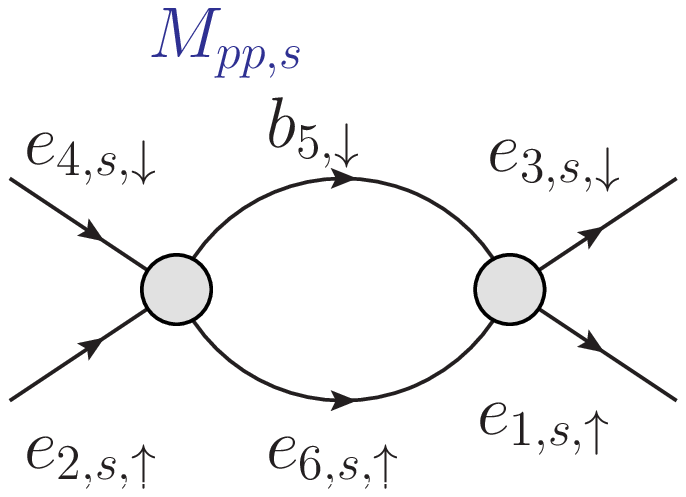}%
\caption{Diagrammatic representations of the processes leading to the bulk-mediated effective two-particle interaction in Eq. (\ref{H_twoParticle}). The diagrams correspond to formulas \eqref{Lph}-\eqref{Mpp}. 
The particle-hole diagrams (index ph) can be split into a direct and a crossed particle-hole channel, where the direct particle-hole channel is only available for equal spins. 
Note that for the interactions of type $M$ one of the intermediate states is an edge state. Due to the sublattice polarization of the edge states, this results in the rule that
all external edge state lines must correspond to the same sublattice and therewith to the same edge.}%
\label{Lph_diagramm}
\end{figure}

The two-particle interactions have the following form:
\begin{widetext}
\begin{equation} 
H_{\rm{SW}}^{2p}  =-U^{2}\sum_{\substack{1,2,3,4\\s,s'}} \left\{ \sum_{\tau, \tau'}\left( L_{\text{ph }s,s'}^{1234}+ \delta_{s,s'} M_{\text{ph }s}^{1234}  \right)
:e_{1s\tau'}^{\dagger}e_{2s'\tau'}e_{3s'\tau}^{\dagger}e_{4s\tau}:
-2 \left(L_{\text{pp }s,s'}^{1234}+\delta_{s,s'} M_{\text{pp }s}^{1234} \right)
:e_{1s\uparrow}^{\dagger}e_{2s'\uparrow}e_{3s\downarrow}^{\dagger}e_{4s'\downarrow}:\right\} \label{H_twoParticle}
\end{equation}
\end{widetext}
where 
\begin{equation} 
L_{\text{ph }s,s'}^{1234} =\sum_{\substack{5\ \rm{occ} \\ 6\rm\ {emp}}}\frac{1}{\epsilon^{b}_{5} - \epsilon^{b}_{6}} \Gamma^{eebb}_{1456,s}\Gamma^{bbee}_{6532,s'}  \label{Lph} 
\end{equation}
and 
\begin{equation}
L_{\text{pp }s,s'}^{1234} =\sum_{\substack{5\ \rm{occ} \\ 6\rm\ {occ}}}\frac{1}{\epsilon^{b}_{5} + \epsilon^{b}_{6}} \Gamma^{ebeb}_{1536,s}\Gamma^{bebe}_{5462,s'}  \label{Lpp}
\end{equation}
are essentially particle-hole and particle-particle response functions, respectively (see Fig. \ref{Lph_diagramm}). 
The summations of $5,6$ go over all bulk states with $\epsilon^b > E_{\rm{Fermi}}$ if they are empty [=emp]  or over all states with $\varepsilon^b< E_{\rm{Fermi}}$
if they are occupied[ =occ], respectively.
In particular, $L^{1221}_{ph}$ can be understood as the bulk-states-mediated RKKY interaction between the spins of the electrons occupying the edge states 1 and 2. 
As expected, this coupling is ferromagnetic between spins on the same edge and antiferromagnetic between opposite edges, which is a consequence of the sublattice polarization of the edge states. 
$L_{ph}$ is one of the most important bulk corrections. However, the particle-particle contribution, which essentially transports a pair of electrons over a wide range of distances, is suppressed by the strong zero'th order on-site Hubbard repulsion, as the latter penalizes occupations different from one electron per state.
The $M$ parameters
\begin{align}
M_{\text{ph }s}^{1234} &=\sum_{5\ \rm{occ}}\sum_{6} \frac{1}{\epsilon^{b}_{5}} \Gamma^{ebee}_{1564,s} \Gamma^{eebe}_{3652,s}  \label{Mph}\\
M_{\text{pp }s}^{1234} &=\sum_{5\ \rm{occ}}\sum_{6} \frac{1}{\epsilon^{b}_{5}} \Gamma^{eeeb}_{1635,s} \Gamma^{beee}_{5462,s} \label{Mpp}
\end{align}
are particle-hole and particle-particle loops where the interaction is mediated by one bulk and one (inactive) edge state (see Fig. \ref{Lph_diagramm}). 
They only give a contribution between edge states living on the same sublattice, and hence on the same side of the ribbon.

Equation \eqref{H_twoParticle} can be further simplified by writing it in terms of spin- and particle-counting operators. For this we distinguish the cases $s=s'$, i.e., intra-edge interactions, and $s\neq s'$, corresponding to spin couplings between opposite edges.
For the latter we obtain
\begin{gather}
 H_{\text{SW}}^{2p, \rm inter} =\frac{U^{2}}{2}\sum_{1234, s}\bigg(\sum_{\mu=0,x,y,z}L_{\text{ph }s,s'}^{1234} \hat{\sigma}^{\mu}_{14,s}\hat{\sigma}^{\mu}_{32,\bar{s}} \nonumber \\
 +4L_{\text{pp }s,\bar{s}}^{1234} :e_{1s\uparrow}^{\dagger}e_{2\bar{s}\uparrow}e_{3s\downarrow}^{\dagger}e_{4\bar{s}\downarrow}:\bigg)\text{ ,}\label{H_2p_inter}
\end{gather}
where
\begin{equation}
\hat\sigma_{12,s}^\mu = \sum_{\tau\tau'} \sigma_{\tau\tau'}^\mu e^\dagger_{1s\tau} e_{2s\tau'},
\end{equation}
and 
where $ \sigma^{x,y,z}$ and $\sigma^0$ are the Pauli and unit matrices, respectively.
The first term in Eq. (\ref{H_2p_inter}) gives rise to an effective spin coupling, whereas the last describes the ``pair hopping'' 
of two electrons from one side to the opposite. Again, this latter process is strongly suppressed by the effective Hubbard interaction.

For the intra-edge case $s=s'$ there are several possible simplifying formulations, depending on which two sites are combined to a spin operator.
One convenient possibility is
\begin{multline}
H_{\text{SW}}^{2p, \rm intra} =\frac{U^{2}}{2} \sum_{\substack{1234\\s}} \biggl\{ A_-^{1234} \sum_{\mu=xyz} \hat\sigma^{\mu}_{12,s} \hat\sigma^{\mu}_{34,s} \\+ A_+^{1234}  \hat\sigma^{0}_{12,s} \hat\sigma^{0}_{34,s} \biggr\},\label{Hferm_intra}
\end{multline}
with
\begin{equation}
A_\pm^{1234} =  \left[L_{{\rm ph},ss}^{1432} + M^{1432}_{{\rm ph},s} \pm \frac{1}{2} (L^{1234}_{{\rm pp},ss}  +M^{1234}_{{\rm pp},s} ) \right].
\end{equation}
It is apparent from Eq. (\ref{Hferm_intra}) that the bulk again mediates a spin-spin interaction between the electrons occupying the edge states.

\subsection{One-particle terms}
\begin{figure}[htb] 
\centering
\includegraphics[width= 0.45\linewidth ]{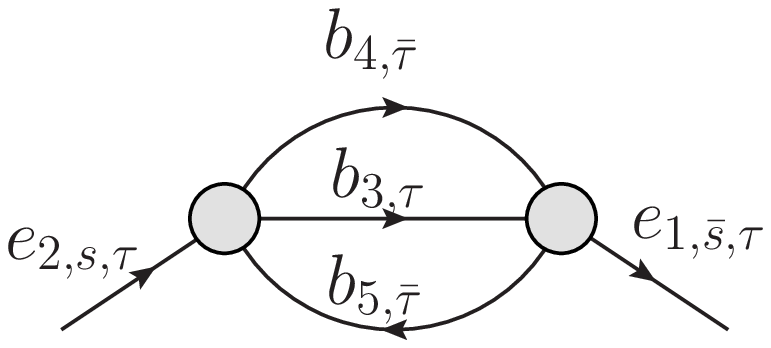}%
\includegraphics[width= 0.45\linewidth]{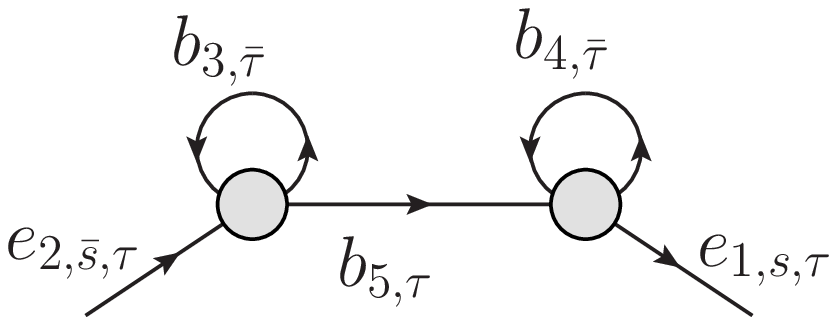}%
\caption{There are two one-particle processes that give rise to an effective hopping from one edge to the other purely mediated by bulk states. 
The processes depicted here have to be summed over the momentum indices as well as over the band indices for the bulk states.}%
\label{B_and_T}
\end{figure}

In addition to the effective two-particle interaction [Eq. (\ref{H_twoParticle})] we obtain bulk-mediated one-particle hopping terms from the Schrieffer-Wolff transformation
\begin{equation}
H_{\text{SW}}^{1p}=-U^{2}\sum_{1,2}\sum_{s,\tau}2B^{1,2}_{s,\bar{s}}:e_{1s\tau}^{\dagger}e_{2\bar{s}\tau}: \label{H_1psw}
\end{equation}
where the hopping amplitude
\begin{equation}\label{B_and_T_formula}
B^{12} =\sum_{\substack{3, 4 \ \rm{emp} \\ 5\ \rm{occ}}}\frac{ \Gamma^{bbeb}_{5314,s}\Gamma^{bebb}_{3245,\bar{s}}}{\epsilon^{b}_{3}+\epsilon^{b}_{4}-\epsilon^{b}_{5}}+ \sum_{\substack{3, 4 \ \rm{occ} \\ 5\ \rm{emp}}}\frac{1}{\epsilon^{b}_{5}} \Gamma^{bbbe}_{3352,\bar{s}} \Gamma^{ebbb}_{1544,s}
\end{equation}
is a sum of two basic processes. Those are visualized diagramatically in Fig. \ref{B_and_T}.

\subsection{Three-particle term}
There is one three particle interaction:
\begin{multline} \label{interactionW}
H_{\text{SW}}^{3p}=-U^{2}\sum_{\substack{1,3,5\text{  emp} \\ 2,4,6\text{ occ}}}\sum_{s,\tau}2W^{\substack{1,2,3 \\ 4,5,6}}_{s,\bar{s}} \\
:e_{1s\bar{\tau}}^{\dagger}e_{2\bar{s}\bar{\tau}}e_{3s\tau}^{\dagger}e_{4\bar{s}\tau}e_{5\bar{s}\tau}^{\dagger}e_{6s\tau}:\text{ ,} 
\end{multline}
where $\tau$ is the spin and the prefactor describes the contraction
\begin{equation} \label{defineW}
W^{\substack{1,2,3 \\ 4,5,6}}_{s,\bar{s}}= \sum_{7\in\mathcal B} \frac{1}{\epsilon^{b}_{7}} \Gamma^{ebee}_{1736,s} \Gamma^{beee}_{7254,\bar{s}} 
\end{equation}
This term combines a hopping of one electron from one edge to the other with a two-particle exchange interaction on different edges.
Since the overlap between different edge states is much smaller than the overlap between an edge state with itself, 
this term should--as long as it carries six different indices--usually be small compared to the other contributions.

\section{Wannier basis\label{sect_wannier_cutoff}}
\subsection{Implementation}
While in the above section the indices ${1,2,3,...}$ describe a general basis, the actual choice of a real basis is important
to specify the wavefunctions and calculate numerical results. 
In the original operators $c_{i\tau}$ [Eq.\eqref{H_original}], the real space multiindex can be split into $i=\left(n,m,s\right)$, where $n,m$ are the coordinates 
along the two lattice basis vectors [cf. Fig. \ref{fig_lattice_bandstructure} (a)] and $s$ is the sublattice index. This gives
\begin{align}
 H_{0} &=-t\sum_{n,m,\tau}c_{nmA\tau}^{\dagger}  \nonumber\\
  &\  \left(c_{nmB\tau}+c_{n(m-1)B\tau}+c_{(n+1)(m-1)B\tau}\right)+\text{H.c.,}
\end{align}
which is conveniently transformed into momentum space along the $n$ direction (parallel to the zigzag edges) followed by a shift $k\rightarrow k+\frac{\pi}{2}$.
With this, the Hamiltonian becomes real, i.e.,
\begin{equation} \label{HamiltonianKspace}
H_{0}=-t\sum_{k,m,\tau} c_{kmA\tau}^{\dagger}\left\{ c_{kmB\tau}+2\sin\frac k2c_{k\left(m-1\right)B\tau}\right\} +\text{H.c.},
\end{equation}
where the basis transformation reads
\begin{equation}
c_{kms\tau} =  \frac{1}{\sqrt{N}}\sum_{n=1}^{N} -e^{i(k-\pi/2)(n-m/2)} c_{nms\tau}.
\end{equation}
In this basis the eigenstates $d^\dagger_{kn\tau}$ with eigenenergy $\epsilon_{kn}$ (i.e., $[H_0, d^\dagger_{kn\tau}] = \epsilon_{kn} d^\dagger_{kn\tau}$) can be written as
\begin{equation}
d^\dagger_{k\nu\tau} = \sum_{m,s} \phi_{k\nu}(m,s) c^\dagger_{kms\tau},\label{kspace_basis}
\end{equation}
where the wave functions $\phi_{k\nu}(s)$ can easily be obtained by diagonalizing the matrix representation of Eq. (\ref{HamiltonianKspace}). 
The subband index $\nu$ labels the different eigenstates corresponding to a given momentum $k$ [see Fig. \ref{fig_lattice_bandstructure} (b)].
For $|k| < k_{\rm lim}$, two of these eigenstates are edge states, for which we use the label $\nu=\pm$ (see also Sec. \ref{sect_first_order}) 
and the operator symbol $e^\dagger_{k\pm,\tau}$. $\pm k_{\rm lim}$ are the limits of the part of the Brillouin zone in which edge states exist. 
Usually, it is close to $\pi/3$. The wave functions of the eigenstates $e^\dagger_{k\pm,\tau}$ are linear combinations of two wave functions that are localized at one edge $s=A,B$. Thus, 
\begin{equation}
e^\dagger_{k,A/B,\tau} = \frac1{\sqrt2}(e^\dagger_{k+\tau} \pm e^\dagger_{k-\tau})\label{edge_operators}
\end{equation}
are edge states which are non-vanishing only at edge $A$ or $B$.

Due to translation symmetry the complete wave functions $\psi_{k\nu}(n,m,s) = N^{-1/2} \phi_{k\nu}(m,s) e^{i k n}$ are plane waves along the ribbon. 
One may expect, however, that for systems in which the interaction is much stronger than the kinetic energy, a more localized basis is better suitable. 
%In the edge state subspace [spanned by the states in Eq. (\ref{edge_operators})], this is exactly the case.
Therefore, our next step is the transformation to a maximally localized Wannier basis.
This transformation shall affect only the edge states, i.e., the states $\nu=\pm$ and $|k|<k_{\rm lim}$. It reads
\begin{equation}\label{Wannier_edge}
e_{xs\tau}^\dagger=\sqrt{\frac 1 K} \sideset{}{'}\sum_{\left|k\right|<k_{\rm{lim}}}e^{ikax} e_{ks\tau}^\dagger
\end{equation}
with $K$ being the number of momenta in the primed sum.
Fig. \ref{fig_lattice_wannier} illustrates the spatial profile of these Wannier edge states.
Since the edge states only exist for part of the one-dimensional Brillouin zone [Fig. \eqref{fig_lattice_bandstructure} (b)],
the momentum summation is limited to $\left|k\right|<k_{\rm{lim}}$. To keep the transformation unitary, the scaling factor $a=\pi/k_{\rm{lim}}$ must be used.

Having defined the basis states of our effective low-energy theory in Eq. (\ref{Wannier_edge}), we are now in a position to actually calculate the specific forms of the bulk corrections [e.g. Eq. (\ref{Lph})]. For computational efficiency, we avail ourselves of the momentum conservation along the edge, thus starting from a representation of the vertex functions $\Gamma$ in the basis given in Eq. (\ref{kspace_basis}).
The momentum conservation provides a restriction to three different momenta, 
and the corresponding coefficient includes only the one-dimensional transverse wavefunctions. For example, the density-density overlap reads
\begin{align}
\Gamma^{eebb,s}_{k,k'-q,(k',b_{1}),(k+q,b_{2})}&=\frac{1}{N}\sum_{m}\phi_{k,s}^{*}\left(m,s\right)\phi_{k'-q, s}\left(m,s\right) \nonumber \\
&\quad \phi_{k',b_{1}}^{*}\left(m\right)\phi_{k+q, b_{2}}\left(m\right) 
\end{align}
where for the bulk states we must choose the band indices $b_{1},b_{2}$. 
With this, all bulk-mediated couplings in the effective theory can be calculated. Here we give only one example, namely the particle-hole loop
\begin{widetext}
\begin{equation}
L_{\text{ph }s,s'}\left(k,k',q\right) =\sum_{\substack{b_{2}\rm{\ occ} \\ b_{3} \rm{\ emp}}}\sum_{p}\frac{\Gamma_{k,k+q,\left(p,b_{2}\right),\left(p-q, b_{3}\right)}^{eebb,s}\Gamma_{\left(p-q,b_{3}\right),\left(p, b_{2}\right),k',k'-q}^{bbee,s'}}{\epsilon_{b_{2},p}-\epsilon_{b_{3}, p-q}}\text{ ,} 
\end{equation}
\end{widetext}
where the summations of $b_{2},b_{3}$ go over all bandindices above or below the Fermi surface, 
depending on whether the state should be empty or occupied.

The general transformation of these coefficients to the Wannier basis is involved but straightforward. Since we are essentially interested 
in cases with two identical indices (compare Sec.\ref{ApproxOneHopping}), and all $\phi_{k\nu}(m,s) \in \mathcal{R}$, this transformation simplifies
to two different cases, namely
\begin{equation}
L_{\text{ph}}^{xxx'x'}=L_{\text{ph}}^{xx'xx'}
= \frac{1}{K^{2}}\sum_{k,k',q}e^{ia(k+k')(x-x')}L_{\text{ph}}\left(k,k',q\right)\text{ ,} 
\end{equation}
which takes a non-vanishing value only for edge states on the same sublattice, and
\begin{equation}
L_{\text{ph}}^{xx'x'x}=\frac{1}{K^{2}}\sum_{k,k',q}e^{ia q(x'-x)}L_{\text{ph}}\left(k,k',q\right)
\end{equation}
which contributes to intra-edge as well as to the inter-edge terms. The remaining interaction terms are calculated in an analogous way. 

The effective hopping is obtained from Eq.\eqref{tStarK} and transformed via
\begin{equation}
t_{xx'}=\frac{1}{K}\sideset{}{'}\sum_{\left|k\right| < k_{\rm{lim}}}t(k)e^{ia(x-x')k} 
\end{equation}
so that the noninteracting pure edge Hamiltonian in the Wannier basis reads
%% WE NEED THIS FORMULA IN THE NEXT CHAPTER
\begin{equation}
H_{0,\rm edge} = \sum_{xx',\tau} t_{xx'} e^\dagger_{xA\tau} e_{x'B\tau} + H.c. \label{h0edge_wannier}
\end{equation}

\subsection{Approximation: Maximally one hoppping} \label{ApproxOneHopping}
In the localized Wannier basis, the term $U\Gamma^{eeee}_{xxxx}$ is larger than all other contributions. Due to this dominant effective Hubbard Hamiltonian,
half filling is strongly favored for the Wannier states, and all processes that map a homogeneous charge occupation in the edge state to a state with broken charge homogeneity will be suppressed.
When gaining an overview over the large variety of interactions in Sec. \ref{sect_ferm_bulk_corr}, we therefore want to focus on those that create and annihilate maximally one doubly occupied or empty state. 
When doing so, we are only left with two sorts of terms: 
\begin{itemize}
 \item There are two-particle terms which have two pairs of equal external indices, they describe a spin-spin interaction.
 \item The second class of terms are those in Eq.\eqref{B_and_T_formula}, describing a single hopping between sites on different sublattices.
 In the final interaction this adds to the effective hopping $t_{xx'}$, but both are suppressed by the onsite repulsion $U^{*}$.
\end{itemize}
The three-particle term \eqref{interactionW} does not contribute at all in this approximation, because it either couples more than two sites, 
or it is multiplied with an expectation value in normal order, which also yields zero
\begin{equation} \label{w_Vanish}
W_{\bar{s}s}(x,y)=\sum_{2,3\in\mathcal{E}}W_{\bar{s}s}^{\substack{x,y,3 \\ 3,2,2}} \left<:\hat{n}_{2\bar{\tau}}:\right>\left<:\hat{n}_{3\bar{\tau}}: \right>=0 
\end{equation}
All two-particle interactions with three different external indices vanish due to the previous approximations: 
They either contain contributions $\propto e_{x,\tau}^{\dagger}e_{z,\tau}e_{y\tau'}^{\dagger}e_{z\tau'}$ (or with two identical creation operator indices, respectively);
this process has a double unbalanced hopping and couples to a double occupied (or empty) state.
Or otherwise interactions of three edge states contain one particle-counting operator, which vanishes analogous to Eq. \eqref{w_Vanish} due to normal ordering.

\section{Corrections to the spin model} \label{sec:SpinModel}

In Section \ref{sect_ferm_bulk_corr} we have derived the bulk corrections to the edge state theory in a completely general fermionic basis. But we have also emphasized that the structure of the corrections becomes clearer if we use a localized basis, namely the Wannier basis we have discussed in the previous section. In the present section we will use this insight for a further approximation, namely the reduction to a spin model for edge magnetism. A similar derivation has been done before in Refs. \onlinecite{golor_quantum_edge_magnetism,schmidt_eff_vs_qmc_2013}, but here we explicitly study the effect of the bulk states, which have been neglected up to now.

It is most important to note that, as stated above, if expressed in the Wannier basis, the largest term in the total Hamiltonian $H_{0, \rm edge} + H_{U,\rm edge} + H_{\rm SW}$ [Eqs. (\ref{h0edge_wannier}), (\ref{HUeeee}), and (\ref{HSW_terms})] is the effective Hubbard interaction
\begin{equation}
U \sum_{x,s} \Gamma_{xxxx} (e^\dagger_{xs\uparrow} e_{xs\uparrow}-1/2)(e^\dagger_{xs\downarrow} e_{xs\downarrow}-1/2).\label{eff_hubbard}
\end{equation}
In the zigzag geometries, $\Gamma_{xxxx}\approx0.1$ is independent of the position of the Wannier state $x$ and only very weakly dependent on width and length of the ribbon. Due to the dominance of the term (\ref{eff_hubbard}), each Wannier edge state is basically occupied by one electron. All other terms affect only the remaining degree of freedom, namely the spin of the localized electrons.

There are two different types of terms in the fermionic Hamiltonian that lead to spin-spin couplings. (a) Terms that can be written as spin-spin interactions in the fermionic basis, such as Eqs. (\ref{H_2p_inter}) and (\ref{Hferm_intra}), and (b) terms that result in a hopping between Wannier states, such as Eqs. (\ref{H_1psw}) and (\ref{h0edge_wannier}).
Fermionic terms of type (a) translate one-to-one to spin interactions. They leave the one-electron-per-Wannier-state subspace invariant. The type-(b) terms do not leave this subspace invariant because they change the occupation of Wannier states. Thus, the corresponding spin couplings are obtained in second order perturbation theory. 

We are interested only in the most important spin-spin interactions. These result from the two-center terms, i.e., the terms in the full fermionic Hamiltonian in which only two different Wannier state positions ($xs$ and $x's' $) appear. These terms have nontrivial actions only on the subspace spanned by the Wannier states of these two sites, so that we may restrict our considerations to a two-site Fock basis $\left\{ \ket{\uparrow\uparrow},\ket{\downarrow\downarrow},\ket{\uparrow\downarrow},\ket{\downarrow\uparrow},\ket{\updownarrow-}\ket{-\updownarrow}\right\} $, where the last two states stand for the double occupation of one of the Wannier states. With a shift in the total energy, all terms in the full fermionic Hamiltonian relevant for these two Wannier states can be written as a $6\times6$ matrix, consisting of $2\times 2$ blocks
\begin{equation}
\bordermatrix{~ &\ket{\uparrow \uparrow}\ket{\downarrow \downarrow} &\ket{\uparrow \downarrow}\ket{\downarrow\uparrow} &\ket{\updownarrow \text{---}} \ket{\text{---} \updownarrow} \cr
\substack{ \ket{\uparrow \uparrow} \\ \ket{\downarrow \downarrow}} &\boldsymbol{0} &\boldsymbol{0} &\boldsymbol{0}  \cr
\substack{ \ket{\uparrow \downarrow} \\ \ket{\downarrow \uparrow}} &\boldsymbol{0}  &H_{\rm{le}} & H_{\rm{hop}}  \cr
\substack{ \ket{\updownarrow \text{---}}  \\ \ket{\text{---} \updownarrow}} &\boldsymbol{0} &H_{\rm{hop}}^{\dagger} &H_{\rm{he}} },\label{blockmatrix6by6}
\end{equation}
where $H_{\rm le}$ contains all terms of type (a), and $H_{\rm hop}$ all terms of type (b). The particular forms of these two blocks depend on the mutual positions of the two Wannier centers under consideration and will be discussed in the following two paragraphs. The block $H_{\rm he}$ contains the effective Hubbard interaction [Eq. (\ref{eff_hubbard})]. It separates the high-energy subspace with doubly occupied Wannier states from the low-energy subspace with single occupations. Due to SU(2) invariance, the states $\ket{\uparrow\uparrow}$ and $\ket{\downarrow\downarrow}$ are decoupled from the other states.

As usual, the residual spin dynamics in the low-energy subspace is then obtained by second order perturbation theory
\begin{equation}
H_{\rm{spin}}=H_{\rm{le}} - H^{\phantom \dagger}_{\rm{hop}} H_{\rm{he}}^{-1} H_{\rm{hop}}^{\dagger}.
\end{equation}
$H_{\rm spin}$ can be expressed in terms of spin operators. In the following we discuss the two different cases of interedge coupling (the Wannier sites are on different edges) and intra-edge coupling (the two Wannier sites are on the same edge $s=s'$).

\subsection{Inter-edge coupling }
We consider two Wannier states $xA$ and $x'B$ on opposite edges.
The low energy part of the inter-edge coupling reads
\begin{equation}
H_{\rm{le}}^{\rm{inter}}= \bordermatrix{ ~ &\ket{\uparrow\downarrow} & \ket{\downarrow\uparrow} \cr \ket{\uparrow\downarrow} & -\chi_{xx'} & \chi _{xx'}
 \cr \ket{\downarrow\uparrow} &\chi_{xx'} &-\chi_{xx'}}  , 
\end{equation}
where
\begin{equation}\label{def_chi}
\chi_{xx'} =2U^{2}L_{\text{ph} AB}^{xx'x'x}
\end{equation}
stems from Eq. (\ref{H_2p_inter}).
The factor 2 occurs because there are two possibilities to assign the indices $x$ and $x'$ to the general indices 1,2,3,4 in \eqref{H_twoParticle}. 
The edge states of opposite edges are restricted to different sublattices, so that the 'direct' two-particle terms from Eq. (\ref{HUeeee}) vanish. 
Thus, there are only the bulk-mediated SW couplings in $H_{\rm le}$. It should be noted that $\chi_{xx'}$ can be interpreted as the bulk-mediated RKKY interaction between the spins of the electrons in the two Wannier states.

The Hamiltonian of the high energy subspace can be well approximated by the effective Hubbard interaction [Eq. (\ref{eff_hubbard})]
\begin{equation}
H_{\rm{he}}^{\rm{inter}}= U \Gamma_{xxxx}  \boldsymbol{1}^{2\times 2},
\end{equation} 
where one should remember that $\Gamma_{xxxx}$ is independent of the Wannier position $x$ for zigzag ribbons.

The block connecting low- and high-energy sector reads
\begin{equation}
H_{\rm hop} = (t_{xx'} + \eta_{xx'}) \begin{pmatrix} 1& 1 \\ -1& -1 \end{pmatrix},
\end{equation}
where $\eta_{xx'} = 2U^{2} B^{xx'}$ is a bulk-mediated one-particle SW hopping [Eq. (\ref{H_1psw})] and $t_{xx'}$ is the direct hopping of the edge states [Eq. (\ref{h0edge_wannier})].

Since $t_{xx'},\eta_{xx'} \ll \Gamma_{xxxx}U$ for reasonably wide ribbons, it is justified to apply perturbation theory as described above,
which results (up to a constant) in a Heisenberg inter-edge spin coupling
\begin{equation} \label{H_inter}
H^{\rm{inter}}_{\rm spin}=\sum_{x,x'}\underbrace{\left\{ 2\chi_{xx'} + \frac{4\left(t_{xx'}+\eta_{xx'} \right)^{2}}{U\Gamma_{xxxx}}\right\}}_{=J^{\rm AFM}_{xx'}} \vec{S}_{xA}\cdot\vec{S}_{x'B}.
\end{equation}
$\vec S_{xs} = (S^x_{xs} ,S^y_{xs} ,S^z_{xs} )$ is the vector of Pauli matrices acting on the spin of the electron in the Wannier state at $x,s$. If the bulk states had been neglected, $\chi_{xx'}$ and $\eta_{xx'}$ would be absent. In this case we recover the inter-edge coupling from Sect. \ref{sect_first_order} [Eq. (\ref{first_order_afm})]. The bulk state contributions $\chi_{xx'}$ and $\eta_{xx'}$ modify the coupling constant $J^{\rm AFM}_{xx'}$.

\begin{figure}[htb] 
\centering
\includegraphics[width= \linewidth]{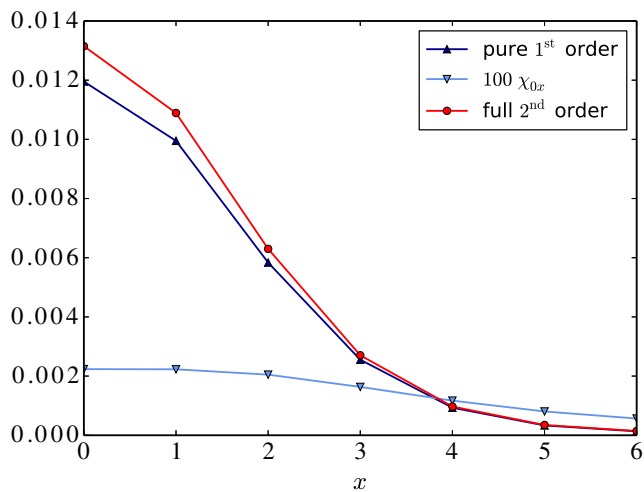}%
\caption{ The different orders of the inter-edge interaction as a function of Wannier index $x$ along the ribbon for $k_{lim}=\pi/3$. The ribbon size is $N=104,\ M=12$ and the original Hamiltonian parameters $U=t=1$.
The direct second-order correction has been scaled to fit in the same plot, and the major contribution to the correction stems from $\eta_{xx'}$.}%
\label{Plot_Inter(n)}
\end{figure}

Figure \ref{Plot_Inter(n)} provides an overview of the different contributions to $J^{\rm AFM}_{xx'}$ as a function of distance $|x-x'|$ between the Wannier states. Clearly, the direct coupling $\sim (t_{xx'})^2/U\Gamma_{xxxx}$ of the edges states is much larger than the bulk-state corrections, which again justifies the use of perturbation theory. Of the bulk state corrections, the bulk-mediated hopping $\eta_{xx'}$ is most relevant. The RKKY-like $\chi_{xx'}$ is only of minor importance.

\begin{figure}[htb] 
\centering
\includegraphics[width= \linewidth]{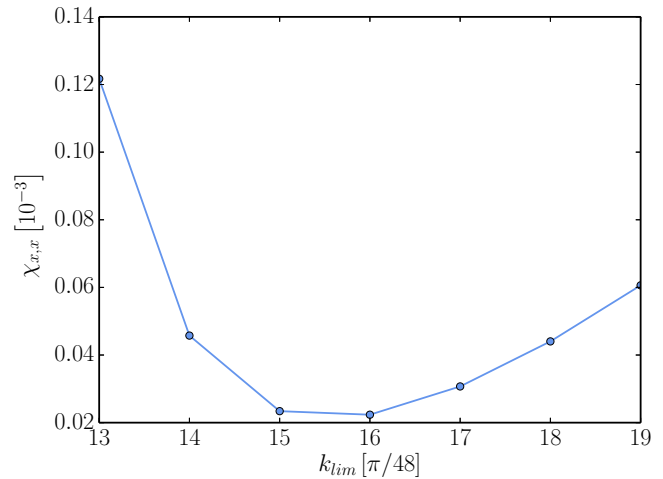}%
\caption{ The direct second order correction term $\chi_{xx}$ (compare \eqref{def_chi}) between directly opposite sites
exhibits a minimum near $k_{\rm{lim}}=\pi/3$. The momentum has been changed in steps of $\delta(k)=\pi/48$. Again the data is $N=104,\  M=12$ and $U=t=1$. }%
\label{Plot_chi(kLim)}
\end{figure}

However, $\chi_{xx'}$ displays another unique feature: Fig. \ref{Plot_chi(kLim)}  shows $\chi_{xx}$ between directly opposite sites as a function of cutoff momentum. 
It has a minimum near $k_{\rm{lim}}=\pi/3$, which justifies this often used choice. The corrections in $\eta_{xx'}$ do not have any special features as functions of cutoff momentum. 
Their values increase monotonically when the cutoff momentum increases, because this leads to a higher degree of delocalization of the edge wavefunctions and thereby to a larger overlap and hopping.

\begin{figure}[htb] 
\centering
\includegraphics[width= \linewidth]{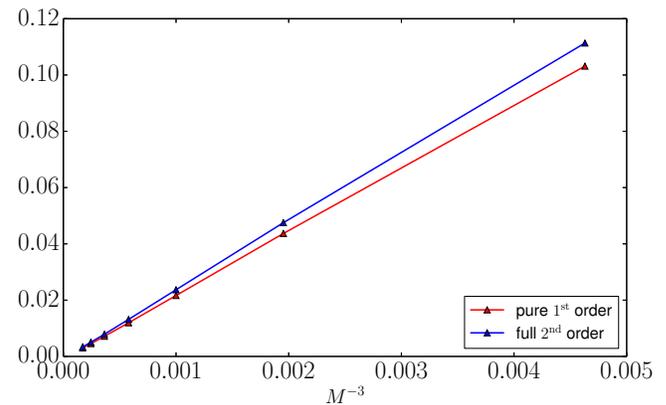}%
\caption{ When the ribbon width $M$  is increased, the zeroth order contribution $J_{0}$ as well as the full term including second order effects decrease as $\frac{1}{M^{3}}$.
This is in accordance with the RKKY interaction in graphene. The further data for this plot is $N=104, k_{\rm{lim}}=\pi/3$ and $U=t=1$ .}%
\label{Plot_allInter(M)}
\end{figure}

Finally, we analyze the dependence of the inter-edge coupling on the width $M$  of the ribbon. In analogy to the RKKY coupling in graphene,\cite{graphene_rkky_brey_fertig}
it is expected to decrease as $\propto M^{-3}$. Figure \ref{Plot_allInter(M)} shows the different orders plotted over this value.
The zeroth order as well as the full expression including second order clearly show the estimated behavior.

\begin{figure}[htb] 
\centering
\includegraphics[width= \linewidth]{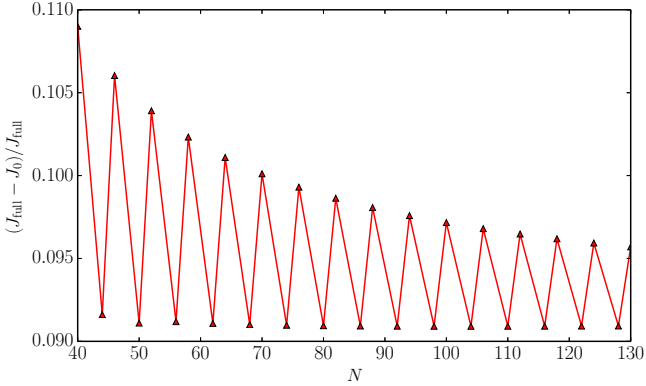}%
\caption{ The relative error of the pure zeroth order calculation  $(J_{0}-J_{\rm{full}})/J_{\rm{full}}$ decreases as a function of ribbon length $N$. 
The zigzag feature of the graph is a finite size effect.
We can deduce that second order correction terms do not become dominant for the inter-edge interaction in long ribbons. The plot shows the results for $M=12, \ k_{\rm{lim}}=\pi/3$ and again $U=t=1$.}%
\label{Plot_error(N)}
\end{figure}

For very broad ribbons, the second order terms consequently do not become dominant over the zeroth oder approximation. Likewise the behavior for constant width $M$, but growing length $N$ shall be checked.
Fig. \ref{Plot_error(N)} shows the relative error of the pure zeroth order inter-edge terms with respect to the terms including the interaction via the bulk, as a function of ribbon length $N$.
One sees for $k_{\rm{lim}}=\pi/3$ a different behavior depending on the value of $N \text{ mod } 3$, which is due to the varying choice of the discrete momenta in the Brillouin zone.
However, for growing ribbon length, this difference diminishes and, which is most important, the overall relative error becomes smaller. Therefore we can conclude that neither for very broad nor for very
long ribbons the bulk interaction terms become dominant, but in both cases the pure first order effective model constitutes a good approximation.

\subsection{Intra-edge coupling }
For the interaction between two states on the same edge---and thus on the same sublattice---first
order spin couplings of the form  $g_{xx'}=\Gamma_{xxx'x'}^{eeee}$ appear in the low-energy part of Eq. (\ref{blockmatrix6by6}), which reads
\begin{equation}
H^{\rm{le}}_{\rm{intra}}= \left(g_{xx'}- \kappa_{xx'} \right) \bordermatrix{ ~ &\ket{\uparrow\downarrow} & \ket{\downarrow\uparrow} \cr \ket{\uparrow\downarrow} &1 & -1 
 \cr \ket{\downarrow\uparrow} &-1 &1 } 
\end{equation}
with 
\begin{equation} \label{kappa}
\kappa_{xx'}=\chi_{xx'}-\theta_{xx'}+\tilde{\chi}_{xx'}-\tilde{\theta}_{xx'} , 
\end{equation}
where $\chi_{xx'}$ has been defined in Eq. \eqref{def_chi}, and
\begin{equation} \label{theta_detail}
 \theta_{xx'} =2U^{2}L_{\rm{ph}\ s,s}^{xxx'x'}=2U^{2}L_{\rm{pp}\ s,s}^{xx'x'x}
\end{equation}
and the parameters with a tilde contain the analogous contribution of the $M$ loops
\begin{align} \label{chi_tildeANDtheta_tilde}
 \tilde{\chi}_{xx'} &=2U^{2}M_{\rm{ph}}^{xx'x'x}=2U^{2} M_{\rm{pp}}^{xx'xx'}  \\ \tilde{\theta}_{xx'} &=2U^{2}M_{\rm{ph}}^{xxx'x'}=2U^{2} M_{\rm{pp}}^{xx'x'x}.
\end{align}

\begin{figure}[htb] 
\centering
\includegraphics[width= \linewidth]{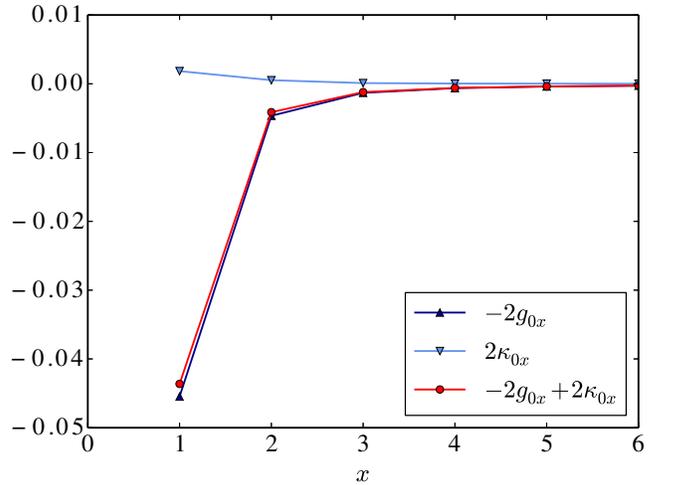}%
\caption{ The intra-edge coupling as a function of distance along the ribbon. The ribbon size is $N=200 \ M=12$ and the Hamiltonian parameters are $U=t=1$, the cutoff momentum is $k_{\rm{lim}}=\pi/3$. 
The pure zeroth order term (dark blue, $\bigtriangleup$) is ferromagnetic, 
the complete second order correction \eqref{kappa} (light blue, $\bigtriangledown$) is antiferromagnetic, so that the full coupling (red, $\circ$) is weakened. 
The correction is about one order of magnitude smaller than the first order.}
\label{Plot_Intra(n)}
\end{figure}

The Hamiltonian of the high energy subspace could again be well approximated by $H_{\rm{he}}^{\rm{intra}}=U^{*}\ \boldsymbol{1}^{2\times 2}$,
but since there is no intra-edge hopping, neither in first nor in second order, there is no need to perform a further perturbation theory.
Thus, the intra-edge spin coupling Hamiltonian, including second order bulk corrections, reads
\begin{equation}\label{H_intra} 
 H^{\rm{intra}}=\sum_{\substack{{x,x'} \\ {x\neq x'}}} -2\left(g-\kappa\right)\vec{S}_{x}\vec{S}_{x' }
\end{equation}
In first order, we have a direct ferromagnetic coupling. Due to the structure of $\kappa_{xx'}$ with positive and negative signs, the effects of different interactions
partially reduce each other.
The term $\chi_{xx'}$, which represents a bulk-states-mediated RKKY coupling, is ferromagnetic, so $\chi_{xx'}<0$, but the overall sum gives $\kappa_{xx'}>0$, so
that the full second order coupling weakens the direct ferromagnetic coupling.
Figure \ref{Plot_Intra(n)}  shows the different contributions as a function of the Wannier index $x$ along the ribbon. 
One sees a rather small reduction of the ferromagnetic coupling when including the Schrieffer-Wolff terms.

\begin{figure}[htb] 
\centering
\includegraphics[width= \linewidth]{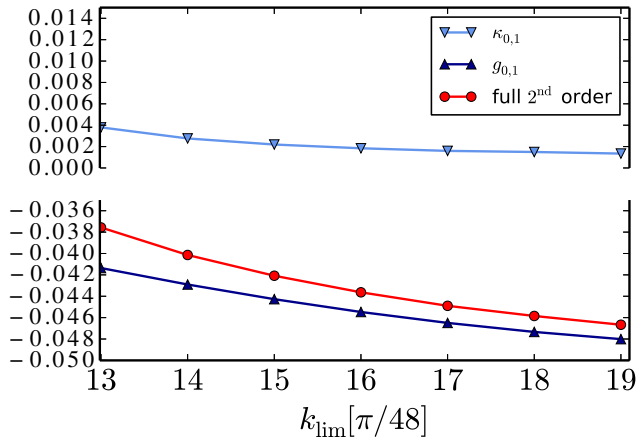}%
\caption{ The first order interaction (dark blue $\bigtriangleup$), the second order interaction (light blue $\bigtriangledown$) and the sum of both (red $\circ$) for nearest neighbors each, 
as a function of cutoff momentum, for $N=200, \ M=12$ and $U=t=1$. The second order only weakly depends on the choice of $k_{\rm{lim}}$.
In the combination, we see no distinctive feature for $k_{\rm{lim}}=\pi/3$, however this can be found for one term of $\kappa_{xx'}$, compare \ref{Plot_chi_Tilde} }%
\label{Plot_Intra(kLim)}
\end{figure}

\begin{figure}[htb] 
\centering
\includegraphics[width= \linewidth]{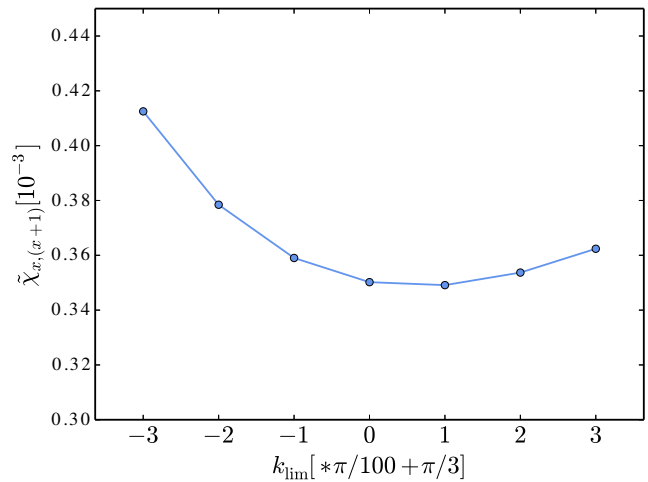}%
\caption{ The interaction summand $\tilde{\chi}_{x,x+1}$ ( compare \eqref{chi_tildeANDtheta_tilde}) between direct neighbors on one edge as it depends on the cutoff momentum,
for $N=200, \ M=12$ and $U=t=1$.
The momentum is varied in steps of $\delta k =\pi / 100$, and the function shows a minimum near  $k_{\rm{lim}}=\pi/3$. 
This summand only yields a small contribution to $\kappa$, but its dependence nonetheless indicates the correct choice of $k_{\rm{lim}}$.} %
\label{Plot_chi_Tilde}
\end{figure}

\begin{figure}[htb] 
\centering
\includegraphics[width= \linewidth]{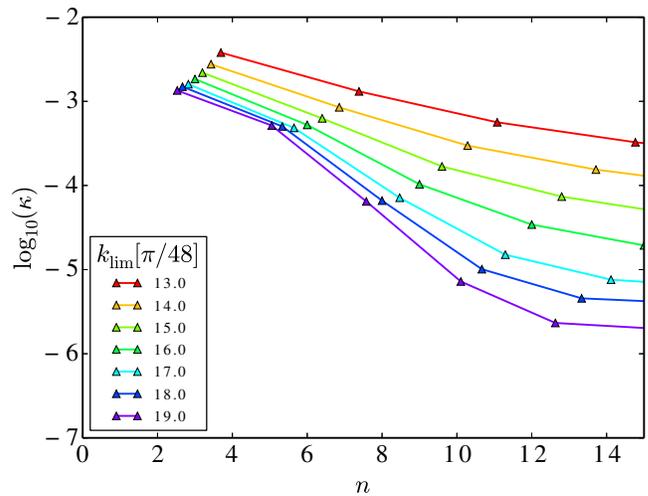}%
\caption{ The logarithm of the pure second-order correction as a function of $n$ gets a kink-like feature, which increases with increasing $k_{\rm{lim}}> \pi/3$, 
and which is caused by $\tilde{\chi}_{xx'}$. Such a feature is probably an artifact and 
we interpret this as a hint for an upper bound in the choice of $k_{\rm{lim}}$. This plot shows the case $N=200, M=12$ and $U=t=1$. } %
%The x-axis is distance along the ribbon coordinate $n$, parallel to the zigazg axis.
\label{Plot_kappa(n)}
\end{figure}

The dependence of the different order terms on the choice of the cutoff momentum is shown in Fig. \ref{Plot_Intra(kLim)}: 
We see that the characteristics of both the first and the second order term are similar, 
and the overall dependence of the full term on the cutoff momentum is only weak.
Figure \ref{Plot_chi_Tilde} shows the dependence of $\tilde{\chi}_{x,x+1}$ on the cutoff momentum;
this is the only intra-edge interaction that shows a feature near $k_{\rm{lim}}=\pi/3$. Even though it only yields a small absolute contribution compared to the other summands in $\kappa_{xx'}$,
this dependence indicates the correct choice of $k_{\rm{lim}}=\pi /3$. The other contributions to $\kappa$ depend monotonically on the cutoff momentum, so that there cannot occur any compensations.
A consequence of this can be seen in Fig. \ref{Plot_kappa(n)}, which shows a logarithmic plot of the pure second-order term as a function of distance $n$, for different values of $k_{\rm{lim}}$. 
We see that for larger $k_{\rm{lim}}$, the graph has an increasing artificially looking kink at the second nearest-neighbor site.

\section{Conclusion}

We have investigated the influence of the bulk states on the magnetic coupling of edge states in graphene nanoribbons with zigzag edges. The edge-bulk interaction has been considered up to second order in a Schrieffer-Wolff transformation. In a first step, all bulk correction terms for an effective model of interacting fermions at the ribbon edges have been derived. Afterwards, this rather complicated fermionic model has been further reduced to a model of effective interacting spins. This model is suitable for large scale simulations. In its simplest form, i.e., without bulk contributions, and for special ribbon geometries it has already been used to study the crossover between quantum- and classical edge magnetism.\cite{golor_quantum_edge_magnetism} One of our main findings in this paper is that the effective spin theory for edge magnetism is applicable not only to the specialized geometries employed in Ref. \onlinecite{golor_quantum_edge_magnetism}, but can be extended to the pure zigzag case, which is in a certain sense the ''worst case'' for these type of effective theories.

Furthermore, we have found that the bulk contributions do not dramatically alter the edge magnetism physics. Hence, the simpler effective model in which the bulk states are ignored  \cite{schmidt_eff_vs_qmc_2013} constitutes a good approximation. If a higher accuracy is required, however, the effective model can be systematically supplemented by bulk corrections via a Schrieffer-Wolff transformation, as shown in this work.
\acknowledgments
We thankfully acknowledge stimulating discussions with M. Golor and S. Wessel. This work was supported by the DFG via Research Training Group 1995 ‘'Quantum many-body methods in condensed matter systems'’ and via the SPP 1459 ''Graphene'' (HO 2422/9-1).

\bibliography{refs}

\end{document}